\documentclass[mathleft]{an}
\usepackage{graphicx}
\usepackage{psfig}
\usepackage{times}
\usepackage{natbib}
\bibpunct{(}{)}{;}{a}{}{,}
\begin{document}


\Yearpublication{}%
\Yearsubmission{2011}%
\Month{}%
\Volume{}%
\Issue{}%

\title{Maximum Reduced Proper Motion Method: Detection of New Nearby Ultracool Dwarfs}

\author{
   N. Phan-Bao\inst{1}\fnmsep\thanks{Corresponding author:
   \email{pbngoc@hcmiu.edu.vn}}
        }
\titlerunning{Maximum Reduced Proper Motion Method}

\authorrunning{N. Phan-Bao}

\institute{Department of Physics, HCMIU, Vietnam National University Administrative Building, Block 6, Linh Trung Ward, Thu Duc District, HCM, Vietnam}

\received{...}
\accepted{...}
\publonline{...}


\abstract{
In this paper, we describe how to use the Maximum Reduced Proper Motion method 
(Phan-Bao et al. 2003) to 
detect 57 nearby L and late-M dwarfs ($d_{\rm{phot}}~\leq~30$~pc): 
36 of them are newly discovered.  
Spectroscopic observations of 43 of the 57 ultracool dwarfs 
were previously reported in Mart\'{\i}n et al. (2010). 
These ultracool dwarfs were 
identified by color criteria in $\sim$5,000 square degrees of the DENIS database
and then further selected by the method for spectroscopic follow-up
to determine their spectral types and spectroscopic distances.
We also report here our newly measured proper motions of these ultracool dwarfs
from multi-epoch images found in public archives (ALADIN, DSS, 
2MASS, DENIS), with at least three distinct epochs and 
time baselines of 2 to 46~years.
}

\maketitle

\section{INTRODUCTION}

Nearby stars are the brightest representatives of their 
class, and therefore provide observational benchmarks
for stellar physics. This is particularly true for 
intrinsically faint objects, such as stars
at the bottom of the main sequence, and brown dwarfs (BDs).

In the last decade, many nearby ($d~\leq~30$~pc) 
ultracool dwarfs ($>\sim$M6) have been discovered
by using the Deep Near-Infrared Survey of the Southern
Sky (DENIS; \citealt{epchtein97}), the Two-Micron All-Sky Survey
(2MASS; \citealt{skrutskie}), the Sloan Digital Sky Survey
(SDSS; \citealt{york} ), the United Kingdom Infrared Telescope
(UKIRT) Infrared Deep Sky Survey (UKIDSS; \citealt{lawrence}).
These ultracool dwarfs have been detected in the
two following ways:
(1) they were purely selected by color-color criteria and then spectroscopic
observations were carried out to determine their ultracool dwarf nature and 
spectral types \citep{del99,martin99,kirk99,cruz03,knapp04,lodieu07};
(2) they are high proper-motion dwarfs (from high proper-motion
catalogs or newly discovered)
and were selected by color criteria, then followed up by spectroscopic
observations \citep*{scholz01,p01,reid02,lepine02,p03,p08}.

In \citet{p03}, we used the Maximum Reduced Proper 
Motion (MRPM) method to select mid-M 
dwarfs (2.0 $\leq$ $I-J$ $\leq$ 3.0, $\sim$M6--M8) in the DENIS
database. This method allowed us to identify not only high-proper
motion (high-pm) M dwarfs but also some with low 
proper motion (low-pm), such as DENIS-P J1538317$-$103850 
(M5, $\mu=20$~mas~yr$^{-1}$). We then extended 
our search over the color range of $3.0 \leq I-J \leq 4.5$ ($\geq$M8)
and successfully applied the method to detect 20 new L and late-M dwarfs 
in the Galactic plane for the southern hemisphere \citep{p08}.
This was the first search of ultracool dwarfs in this crowded area.
The results demonstrate the success of the method, making it a useful tool
for hunting ultracool dwarfs in the solar neighborhood.

Here we applied the same 
technique to identify dwarfs cooler than M8
($I-J \geq 3.0$) at high Galactic latitude in $\sim$5,000 square degrees of the DENIS
database. Fifty seven nearby L and late-M dwarfs
were identified by the MRPM method: 36 of them are new, 21 were previously
reported in the literature. In this search, we additionally revealed
a young ultracool dwarf candidate in the R-CrA molecular cloud region.
\citet{martin10} presented spectroscopic observations of 43 of the 57 nearby 
ultracool dwarfs as well as of the new member candidate of R-CrA
identified in this paper.

We present selection criteria in 
\S~2. We describe
the proper motion measurements in \S~3, and the MRPM filtering
of the candidates in \S~4. \S~5 examines the contamination 
by distant red stars, and discusses some individual interesting
objects. We summarize in \S~6.

\begin{figure}
\psfig{width=8.2cm,file=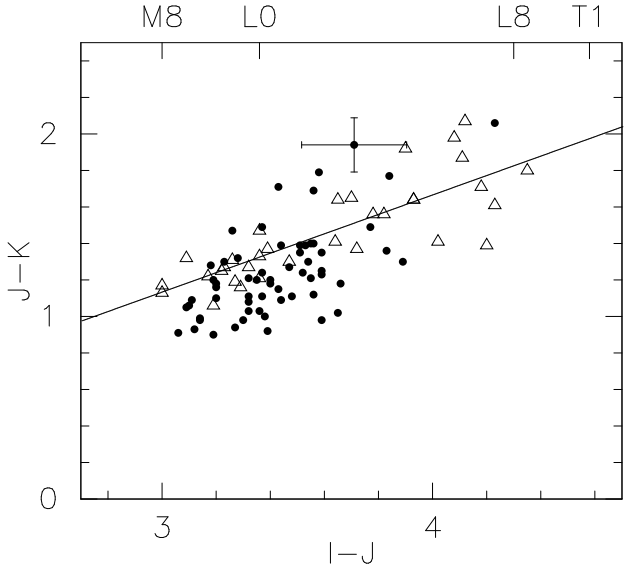,angle=-90}
\caption{($I-J$, $J-K$) color-color diagram for our 60 late-M 
and L dwarf candidates (Table~\ref{candidates}), plotted as 
solid circles, as well as known late-M and L dwarfs 
(see Table~1 in \citealt{p08}; plotted as open triangles).
The line represents our linear fit to the colors of 
the literature dwarfs: $J-K = -0.462 + 0.532(I-J), \sigma=0.15$.  
\label{IJ_JK}}
\end{figure}
\section{SELECTION CRITERIA}
We conducted a systematic search in 
the DENIS database (available at 
the Paris Data Analysis Center, PDAC) for potential 
members of the solar neighbourhood, with simple and 
well defined criteria. We first selected all DENIS 
objects which matched $|b_{II}|~{\geq}~15\degr$ (high 
Galactic latitude) and $I-J \geq 3.0$ (expected spectral type 
later than M8, see \citealt{leggett92}).
We then required that the position of the candidates 
in the ($I-J$, $J-K$) color-color diagram 
be within $J-K = \pm0.5$~mag of our linear fit 
to the locus of ultracool dwarfs (Fig.~\ref{IJ_JK}).

At this point we filtered out all candidates within 
the boundaries of the Magellanic Clouds \citep{cioni} 
and of known high-latitude molecular clouds 
(see Table~2 of \citealt{cruz03} and references therein). 
This eliminated most red giants and reddened blue stars,
at the obvious cost of losing the cool dwarfs which happen
to overlap a cloud. We then computed photometric distances 
from the following ($I-J$, $M_{\rm J}$) relation as given in \citet{p08}
and retained the candidates with $d_{\rm phot} \leq 30$~pc:

\begin{eqnarray}
 M_{\rm J} & = & a_{0}+a_{1}(I-J)+a_{2}(I-J)^{2}+a_{3}(I-J)^{3} \nonumber \\ 
           &   & +~a_{4}(I-J)^{4} \label{eq1}
\end{eqnarray}
where $a_{0}=130.164$, $a_{1}=-124.899$, 
$a_{2}=46.948$, $a_{3}=-7.4842$, $a_{4}=0.4341$, 
valid for $3.0~\leq~I-J~\leq~6.0$. The rms dispersion
around that fit is 0.30~mag, corresponding to a 14\% error 
on distances. 

Before measuring proper motions, which is needed for the use of the MRPM method
to discriminate between nearby dwarfs and distant 
giants, we cross-identified the bright candidates 
($I<13.0$, $d_{\rm phot}<<30$ pc) with both SIMBAD 
and the catalogs of red variable stars (e.g. 
\citealt{pojmanski,samus}) in VIZIER. As expected,
all bright candidates are previously recognized giants. 
This left 60 ultracool dwarf candidates 
(Table~\ref{candidates}) for which we needed to 
measure proper motions. 

\begin{table*}
  \caption{60 DENIS red objects}
  \label{candidates}
  $$
  \begin{tabular}{llllllllll}
   \hline 
   \hline
   \noalign{\smallskip}
 DENIS name          &  $\alpha_{2000}$ &  $\delta_{2000}$ &  Epoch           &
 $I$                 &  $I-J$           &  $J-K$           &  err$I$          &
 err$J$              &  err$K$              \\
   \hline
   \noalign{\smallskip}  
J0000286$-$124514 & 00 00 28.69 & $-$12 45 14.7 & 1998.564 &  16.16 & 3.09 & 1.05 & 0.08 & 0.11 & 0.11 \\
J0006579$-$643654 & 00 06 57.92 & $-$64 36 54.2 & 1998.860 &  16.71 & 3.28 & 1.32 & 0.12 & 0.09 & 0.11 \\
J0014554$-$484417 & 00 14 55.44 & $-$48 44 17.1 & 1996.669 &  17.54 & 3.56 & 1.40 & 0.13 & 0.08 & 0.13 \\
J0028554$-$192716 & 00 28 55.44 & $-$19 27 16.6 & 1998.784 &  17.61 & 3.66 & 1.18 & 0.18 & 0.18 & 0.17 \\
J0031192$-$384035 & 00 31 19.28 & $-$38 40 35.8 & 1999.748 &  17.62 & 3.52 & 1.24 & 0.14 & 0.10 & 0.15 \\
J0050244$-$153818 & 00 50 24.41 & $-$15 38 18.9 & 1999.715 &  16.86 & 3.20 & 1.10 & 0.10 & 0.10 & 0.11 \\
J0053189$-$363110 & 00 53 18.97 & $-$36 31 10.5 & 2000.885 &  18.10 & 3.89 & 1.30 & 0.19 & 0.12 & 0.16 \\
J0055005$-$545026 & 00 55 00.53 & $-$54 50 26.3 & 2000.896 &  17.12 & 3.38 & 1.00 & 0.11 & 0.20 & 0.15 \\
J0116529$-$645557 & 01 16 52.90 & $-$64 55 57.1 & 1996.661 &  17.90 & 3.46 & 1.28 & 0.16 & 0.10 & 0.17 \\
J0128266$-$554534 & 01 28 26.64 & $-$55 45 34.4 & 2000.863 &  17.07 & 3.26 & 1.47 & 0.16 & 0.12 & 0.13 \\
J0147327$-$495448 & 01 47 32.77 & $-$49 54 48.0 & 2000.885 &  16.05 & 3.14 & 0.97 & 0.07 & 0.09 & 0.07 \\
J0206566$-$073519 & 02 06 56.68 & $-$07 35 19.8 & 1999.907 &  17.92 & 3.58 & 1.35 & 0.17 & 0.11 & 0.15 \\
J0213371$-$134322 & 02 13 37.12 & $-$13 43 22.3 & 1998.808 &  17.64 & 3.36 & 1.03 & 0.17 & 0.16 & 0.18 \\
J0227102$-$162446 & 02 27 10.28 & $-$16 24 46.8 & 1998.805 &  17.04 & 3.37 & 1.49 & 0.11 & 0.12 & 0.18 \\
J0230450$-$095305 & 02 30 45.01 & $-$09 53 05.1 & 2000.831 &  18.24 & 3.56 & 1.69 & 0.21 & 0.18 & 0.15 \\
J0240121$-$530552 & 02 40 12.11 & $-$53 05 52.5 & 1998.932 &  18.18 & 3.83 & 1.36 & 0.22 & 0.10 & 0.16 \\
J0254058$-$193452 & 02 54 05.81 & $-$19 34 52.3 & 2000.904 &  16.30 & 3.19 & 1.20 & 0.08 & 0.08 & 0.10 \\
J0301488$-$590302 & 03 01 48.84 & $-$59 03 02.3 & 1996.817 &  16.80 & 3.37 & 1.11 & 0.10 & 0.08 & 0.09 \\
J0304322$-$243513 & 03 04 32.28 & $-$24 35 13.7 & 1996.934 &  17.29 & 3.32 & 1.03 & 0.12 & 0.09 & 0.17 \\
J0325293$-$431229 & 03 25 29.37 & $-$43 12 29.8 & 2000.743 &  17.48 & 3.32 & 1.21 & 0.15 & 0.10 & 0.16 \\
J0407089$-$234829 & 04 07 08.91 & $-$23 48 29.9 & 1998.742 &  16.80 & 3.20 & 1.18 & 0.09 & 0.15 & 0.17 \\
J0419556$-$402534 & 04 19 55.65 & $-$40 25 34.3 & 2000.888 &  17.34 & 3.41 & 1.20 & 0.11 & 0.10 & 0.13 \\
J0427270$-$112713 & 04 27 27.09 & $-$11 27 13.8 & 1996.948 &  16.67 & 3.14 & 0.99 & 0.13 & 0.08 & 0.12 \\
J0436205$-$421851 & 04 36 20.53 & $-$42 18 51.7 & 2000.880 &  18.00 & 3.55 & 1.30 & 0.17 & 0.12 & 0.16 \\
J0501005$-$770549 & 05 01 00.53 & $-$77 05 49.0 & 1999.795 &  17.16 & 3.27 & 0.95 & 0.11 & 0.10 & 0.15 \\
J0526434$-$445544 & 05 26 43.47 & $-$44 55 44.7 & 1996.954 &  17.62 & 3.59 & 1.25 & 0.14 & 0.10 & 0.14 \\
J0605019$-$234227 & 06 05 01.93 & $-$23 42 27.2 & 1999.129 &  17.89 & 3.51 & 1.35 & 0.14 & 0.09 & 0.15 \\
J0624458$-$452156 & 06 24 45.87 & $-$45 21 56.1 & 1997.082 &  18.32 & 3.84 & 1.77 & 0.19 & 0.11 & 0.13 \\
J0627401$-$391933 & 06 27 40.17 & $-$39 19 33.2 & 2000.893 &  17.50 & 3.41 & 1.18 & 0.13 & 0.11 & 0.15 \\
J0641184$-$432233 & 06 41 18.41 & $-$43 22 33.3 & 1998.959 &  16.97 & 3.23 & 1.31 & 0.08 & 0.08 & 0.11 \\
J0703238$-$611004 & 07 03 23.84 & $-$61 10 04.7 & 1996.101 &  17.91 & 3.44 & 1.39 & 0.19 & 0.11 & 0.16 \\
J0719317$-$505141 & 07 19 31.76 & $-$50 51 41.4 & 1996.117 &  17.44 & 3.44 & 1.09 & 0.11 & 0.09 & 0.14 \\
J0921141$-$210445 & 09 21 14.10 & $-$21 04 45.5 & 2001.132 &  16.50 & 3.65 & 1.02 & 0.09 & 0.08 & 0.10 \\
J1019245$-$270717 & 10 19 24.58 & $-$27 07 17.5 & 1997.189 &  16.90 & 3.33 & 1.08 & 0.10 & 0.08 & 0.15 \\
J1115297$-$242934 & 11 15 29.72 & $-$24 29 34.8 & 1998.222 &  16.50 & 3.12 & 0.93 & 0.08 & 0.07 & 0.15 \\
J1206501$-$393725 & 12 06 50.12 & $-$39 37 25.9 & 1996.109 &  17.67 & 3.36 & 1.19 & 0.16 & 0.10 & 0.16 \\
J1234018$-$112407 & 12 34 01.89 & $-$11 24 07.2 & 1999.463 &  18.22 & 3.59 & 1.23 & 0.19 & 0.13 & 0.20 \\
J1256569$+$014616 & 12 56 56.94 & $+$01 46 16.9 & 1996.372 &  18.20 & 3.77 & 1.49 & 0.17 & 0.09 & 0.13 \\
J1359551$-$403456 & 13 59 55.11 & $-$40 34 56.5 & 1996.393 &  16.98 & 3.20 & 1.16 & 0.10 & 0.10 & 0.13 \\
J1411051$-$791536 & 14 11 05.16 & $-$79 15 36.0 & 1996.366 &  16.23 & 3.10 & 1.06 & 0.07 & 0.08 & 0.10 \\
J1446229$-$395231 & 14 46 22.96 & $-$39 52 31.3 & 1998.540 &  17.67 & 3.37 & 1.23 & 0.17 & 0.14 & 0.16 \\
J1520174$-$175530 & 15 20 17.46 & $-$17 55 30.5 & 2000.566 &  17.80 & 3.40 & 0.92 & 0.16 & 0.15 & 0.21 \\
J1622326$-$120719 & 16 22 32.68 & $-$12 07 19.0 & 1997.263 &  16.56 & 3.20 & 0.89 & 0.07 & 0.08 & 0.13 \\
J1633131$-$755322 & 16 33 13.13 & $-$75 53 22.8 & 1999.633 &  16.20 & 3.10 & 1.10 & 0.06 & 0.07 & 0.10 \\
J1647383$-$115608 & 16 47 38.33 & $-$11 56 08.9 & 1999.584 &  18.08 & 4.23 & 2.05 & 0.17 & 0.10 & 0.09 \\
J1707252$-$013809 & 17 07 25.27 & $-$01 38 09.2 & 2000.566 &  17.81 & 3.55 & 1.40 & 0.15 & 0.12 & 0.14 \\
J1716352$-$031542 & 17 16 35.23 & $-$03 15 42.5 & 2001.378 &  14.46 & 3.43 & 1.71 & 0.05 & 0.10 & 0.09 \\
J1753452$-$655955 & 17 53 45.25 & $-$65 59 55.1 & 1996.667 &  17.80 & 3.59 & 1.79 & 0.14 & 0.10 & 0.12 \\
J1901391$-$370017 & 19 01 39.11 & $-$37 00 17.5 & 1999.433 &  17.94 & 3.71 & 1.94 & 0.16 & 0.11 & 0.10 \\
J1907440$-$282420 & 19 07 44.01 & $-$28 24 20.3 & 1999.748 &  17.95 & 3.60 & 0.97 & 0.16 & 0.12 & 0.18 \\
J1934511$-$184134 & 19 34 51.19 & $-$18 41 34.8 & 2000.568 &  17.71 & 3.43 & 1.15 & 0.14 & 0.11 & 0.16 \\
J2013108$-$124244 & 20 13 10.83 & $-$12 42 44.8 & 1998.584 &  18.07 & 3.55 & 1.21 & 0.17 & 0.15 & 0.17 \\
J2126340$-$314322 & 21 26 34.00 & $-$31 43 22.1 & 1998.652 &  16.26 & 3.06 & 0.91 & 0.07 & 0.13 & 0.16 \\
J2139136$-$352950 & 21 39 13.65 & $-$35 29 50.6 & 1996.691 &  17.94 & 3.47 & 1.11 & 0.17 & 0.11 & 0.20 \\
J2143510$-$833712 & 21 43 51.04 & $-$83 37 12.5 & 1998.529 &  16.50 & 3.30 & 0.98 & 0.10 & 0.06 & 0.11 \\
J2150133$-$661036 & 21 50 13.30 & $-$66 10 36.8 & 1998.608 &  17.23 & 3.55 & 1.13 & 0.14 & 0.08 & 0.12 \\
J2150149$-$752035 & 21 50 14.97 & $-$75 20 35.8 & 1998.518 &  17.45 & 3.51 & 1.39 & 0.16 & 0.09 & 0.12 \\
J2243169$-$593219 & 22 43 16.99 & $-$59 32 19.8 & 1998.841 &  17.40 & 3.32 & 1.11 & 0.15 & 0.10 & 0.14 \\
J2308113$-$272200 & 23 08 11.30 & $-$27 22 00.6 & 1998.789 &  18.11 & 3.53 & 1.40 & 0.20 & 0.17 & 0.17 \\
J2345390$+$005514 & 23 45 39.05 & $+$00 55 14.4 & 1998.718 &  16.90 & 3.18 & 1.28 & 0.12 & 0.13 & 0.13 \\
    \noalign{\smallskip}
    \hline 
   \end{tabular}
  $$
\end{table*}

\section{PROPER MOTION MEASUREMENTS}
We queried ALADIN 
\footnote{http://aladin.u-strasbg.fr/java/nph-aladin.pl}
and the Digital Sky Survey (DSS)
server\footnote{http://archive.stsci.edu/cgi-bin/dss\_plate\_finder} 
for publically available scanned plates containing 
these candidates. Both ALADIN and the DSS archive provide
access to the plates of the POSS-I (R-band), SERC-R 
(R-band), SERC-I (I-band), POSS-II F (R-band), and POSS-II N 
(I-band) surveys. ALADIN additionally contains digitized images
of the ESO-R survey (R-band), which often extend the 
time baseline enough to significantly reduce the standard error 
of our proper motion measurements. Forty candidates were detected in R-band. 
We used {\small SEXTRACTOR} 
\citep{bertin} to extract coordinates of our targets from 
all available digitized images. We then determined proper 
motions through a least-square fit to the positions at the three
to five available epochs with time baselines spanning 2--46 yr, 
including DENIS and 2MASS\footnote{http://cdsweb.u-strasbg.fr/viz-bin/VizieR}.
A few objects with short baselines (e.g., DENIS 0230$-$0953, 
Table~2) have large errors. New images for those
would greatly extend their epoch coverage and would pinpoint
their proper motion. Most of the measurements are of much
better quality however, and the lowest confidently measured
proper motions are $\sim$40~mas~yr$^{-1}$ (DENIS 0407$-$2348, 
DENIS 0605$-$2342).

Table~2 and \ref{redobj} list these 
proper motions and the associated standard errors. These tables
include 38 objects with $\mu \geq 0.1\arcsec$yr$^{-1}$ and 
22 with $\mu \le 0.1\arcsec$yr$^{-1}$. Eight of the 38 objects were 
previously reported as high-pm ultracool dwarfs in \citet{deacon}.
\begin{figure}
\psfig{width=8.2cm,file=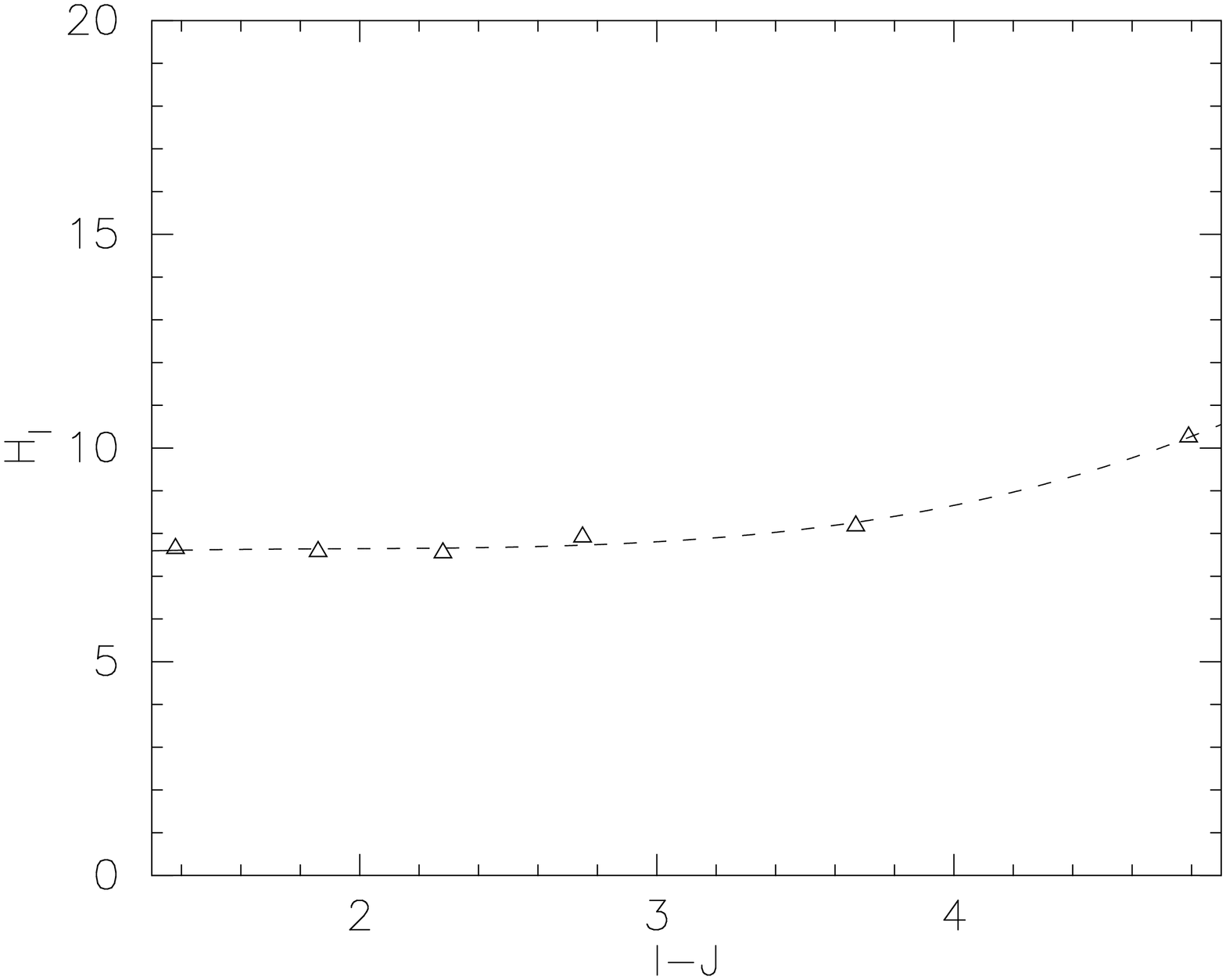,angle=-90}
\caption{($H_{\rm I}^{\rm max}$, $I-J$) $I$-band maximum reduced proper motions-color
diagram for red giants, plotted as open triangles, data from \citet{bessell88,the,fluks}.
The dashed curve represents our cubic least-square fit to the data.
\label{HI_IJ_non_obj}}
\end{figure}
\begin{figure*}
\psfig{width=17.5cm,file=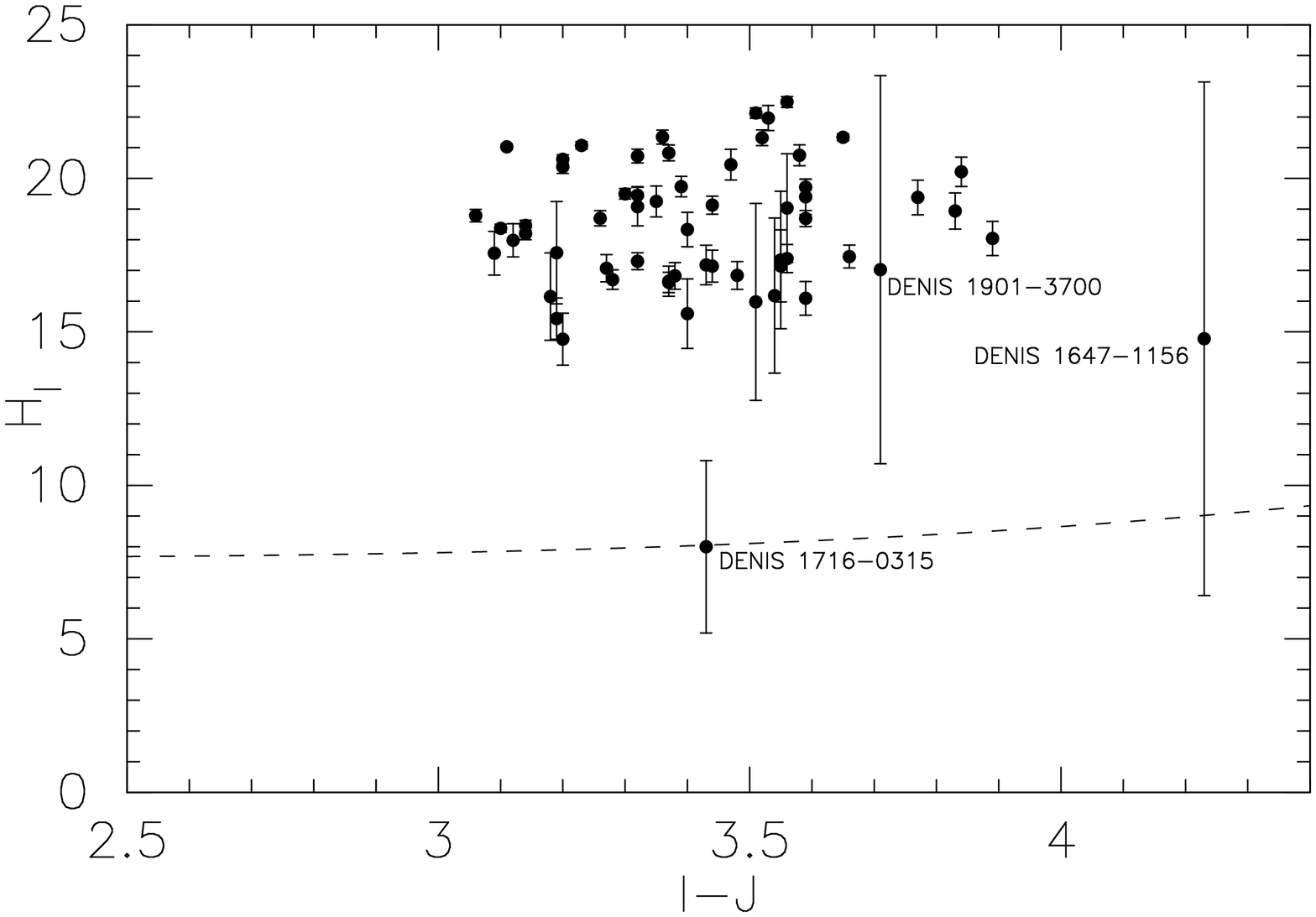,angle=-90}
\caption{$I$-band reduced proper motions versus $I-J$. The dashed 
curve represents the maximum possible reduced proper motion for
a giant at a given color, $H_{\rm I}^{\rm max}$. Objects well above this curve
(e.g., $>3\sigma$) must be
dwarfs. 
Two objects classified as red, distant stars are indicated:
DENIS~1647$-$1156 and DENIS~1716$-$0315. 
DENIS~1901$-$3700 is above the curve but within $2\sigma$, 
see \S~5 for further discussion.
\label{HI_IJ}}
\end{figure*}
\section{L AND LATE-M DWARF CANDIDATES SELECTED BY MAXIMUM 
REDUCED PROPER MOTION}

As in our search for nearby mid-M dwarfs \citep{p03}, 
we use the MRPM method to reject giants. This method uses the 
reduced proper motion ($H=M+5$$\log(V_{\rm t}$ $/4.74)$ = 
$m+5+\log({\mu})$)
versus color diagram, where nearby ultracool dwarfs and distant 
giants segregate very cleanly. We use the 
color-magnitude relation of giants to compute the 
maximum reduced proper motion that a red giant can have at 
a given color by setting its tangential velocity to the
Galactic escape velocity, for which we adopt a conservative 
800~km~s$^{-1}$. We then declare any object with a reduced proper
motion well above ($>3\sigma$) that value for its color to be a dwarf.

To use this method here for the cooler dwarfs (M8--L8), we 
had to extend the maximum reduced proper motion 
curve of giants towards redder colors ($I-J \geq 3.0$), 
adding cooler giants from \citet{fluks} to the
\citet{the,bessell88} sample used in \citet{p03}. 
The following cubic least-square fit (see 
Fig.~\ref{HI_IJ_non_obj}) is valid for 
$3.0 \leq I-J \leq 4.5$, or dwarf spectral 
types between M8.0 and L8.0:

\begin{eqnarray}
   H_{\rm I}^{\rm max} & = & 6.79 + 1.25(I-J)-0.630(I-J)^{2} \nonumber \\
                 &   & +0.108(I-J)^{3} \label{eq2}                 
\end{eqnarray}
The rms dispersion around the fit is 0.1 mag.

Figure~\ref{HI_IJ} shows the position of the resulting 
$H_{\rm I}^{\rm max}$ vs. ($I-J$) curve relative to
the 60 candidates. The curve classifies 57 candidates as
nearby late-M and L dwarfs (Table~2). DENIS~1901$-$3700 
(further discussed in \S~5) is in the dwarf part of the
diagram but within 2$\sigma$ of the limit. 
The remaining two (see Table~\ref{redobj}) are within 1$\sigma$,
they are therefore likely red, distant objects.
Nineteen of the 57 ultracool dwarfs have proper motions
under $0.1\arcsec$yr$^{-1}$, giving a significant $\sim$33\%
fraction of low-pm ultracool dwarfs in our sample. 
\begin{table*}
\rotatebox{90}{
 \begin{tabular}{lcrcrlrlllllllrl}
\multicolumn{16}{p{22.5cm}}{{\bf Table 2.} 57 nearby L, and late-M dwarfs and a member candidate
of R-CrA (DENIS 1901$-$3700)}\\
   \hline 
   \hline
   \noalign{\smallskip}
 DENIS name  &  Other name &  Time &  No. of   &$\mu$RA  &  err$\mu$RA &  $\mu$DE &  err$\mu$DE & 
 $H_{\rm I}$ &  err$H_{\rm I}$ &  $H_{\rm I}^{\rm max}$  & $d$         & err$d$   &  SpT  &  V$_{\rm t}$ &  Refs.  \\
             &             &  b.l. (yr) &  obs.  & (\arcsec yr$^{-1}$) &  (\arcsec yr$^{-1}$) &  (\arcsec yr$^{-1}$) & (\arcsec yr$^{-1}$) &
 (mag) &  (mag)  &  (mag) & (pc)  & (pc) &      &  km s$^{-1}$         &   \\
 (1) & (2) & (3) & (4) & (5) & (6) & (7) & (8) & (9) & (10) & (11) & (12) & (13) & (14) & (15) & (16) \\
   \hline
   \noalign{\smallskip}  
D* 0000$-$1245 & 2M 0000$-$1245  &  5.066 & 5 & $-$0.143 & 0.050 & $-$0.126 & 0.027 & 17.6 & 0.7 &  7.8 & 23.1 & 4.4 & M9.0 & 20.9 &  5    \\
D* 0006$-$6436 &                 &  6.941 & 3 &    0.094 & 0.005 & $-$0.033 & 0.013 & 16.7 & 0.3 &  7.9 & 21.9 & 3.9 & L0.0 & 10.3 &       \\
D* 0014$-$4844 &                 & 13.231 & 4 &    0.936 & 0.013 &    0.278 & 0.030 & 22.5 & 0.2 &  8.1 & 19.1 & 3.3 & L2.0 & 88.4 &       \\
D* 0028$-$1927 &                 & 15.022 & 4 &    0.093 & 0.008 &    0.002 & 0.016 & 17.5 & 0.4 &  8.2 & 16.3 & 3.6 & L3.0 &  7.2 &       \\
D* 0031$-$3840 & S* 0031$-$3840  & 17.912 & 5 &    0.542 & 0.027 & $-$0.098 & 0.015 & 21.3 & 0.3 &  8.1 & 21.4 & 3.9 & L2.0 & 55.9 &  1    \\
D* 0050$-$1538 & S* 0050$-$1538  & 16.044 & 5 & $-$0.243 & 0.022 & $-$0.443 & 0.019 & 20.4 & 0.2 &  7.9 & 26.9 & 5.0 & M9.0 & 64.4 &  1,5  \\
D* 0053$-$3631 &                 & 13.345 & 4 &    0.032 & 0.007 & $-$0.092 & 0.015 & 18.0 & 0.6 &  8.5 & 13.4 & 2.6 & L4.0 &  6.2 &       \\
D* 0055$-$5450 &                 & 19.096 & 5 & $-$0.077 & 0.011 &    0.041 & 0.007 & 16.8 & 0.4 &  8.0 & 22.1 & 5.1 & L1.0 &  9.1 &       \\
D* 0116$-$6455 &                 &  6.038 & 4 & $-$0.171 & 0.013 & $-$0.274 & 0.052 & 20.4 & 0.5 &  8.1 & 26.7 & 4.9 & L1.0 & 40.9 &       \\
D* 0128$-$5545 & 2M 0128$-$5545  & 15.968 & 4 & $-$0.210 & 0.007 &    0.028 & 0.013 & 18.7 & 0.2 &  7.9 & 26.8 & 5.2 & L0.0 & 26.9 &  2    \\
D* 0147$-$4954 & 2M 0147$-$4954  & 21.951 & 5 & $-$0.065 & 0.008 & $-$0.261 & 0.014 & 18.2 & 0.2 &  7.8 & 20.4 & 3.7 & M9.0 & 26.0 &  3    \\
D* 0206$-$0735 &                 & 11.082 & 3 & $-$0.043 & 0.006 &    0.003 & 0.022 & 16.1 & 0.5 &  8.2 & 21.5 & 4.1 & L2.0 &  4.4 &       \\
D* 0213$-$1343 &                 & 11.956 & 3 & $-$0.286 & 0.020 & $-$0.471 & 0.005 & 21.3 & 0.2 &  8.0 & 29.1 & 6.2 & L1.0 & 76.0 &       \\
D* 0227$-$1624 & S* 0227$-$1624  &  8.836 & 4 &    0.472 & 0.023 & $-$0.324 & 0.035 & 20.8 & 0.3 &  8.0 & 21.7 & 4.2 & L1.0 & 58.9 &  1    \\
D* 0230$-$0953 &                 &  2.056 & 3 &    0.144 & 0.103 & $-$0.002 & 0.035 & 19.0 & 1.8 &  8.1 & 26.3 & 5.8 & L2.0 & 18.0 &       \\
D* 0240$-$5305 &                 &  5.824 & 3 &    0.138 & 0.022 &    0.032 & 0.011 & 18.9 & 0.6 &  8.4 & 15.4 & 2.8 & L4.0 & 10.3 &       \\
D* 0254$-$1934 & 2M 0254$-$1934  &  3.246 & 5 &    0.180 & 0.127 & $-$0.008 & 0.102 & 17.6 & 1.7 &  7.9 & 21.1 & 3.7 & M9.0 & 18.0 &  7   \\
D* 0301$-$5903 &                 & 15.899 & 4 &    0.088 & 0.014 & $-$0.031 & 0.011 & 16.6 & 0.5 &  8.0 & 19.4 & 3.4 & L1.0 &  8.6 &  4    \\
D* 0304$-$2435 &                 & 13.039 & 4 &    0.010 & 0.023 & $-$0.100 & 0.005 & 17.3 & 0.3 &  7.9 & 26.7 & 4.8 & L0.0 & 12.7 &       \\
D* 0325$-$4312 &                 &  5.883 & 4 &    0.085 & 0.023 & $-$0.191 & 0.040 & 19.1 & 0.6 &  7.9 & 29.1 & 5.4 & L0.0 & 28.8 &       \\
D* 0407$-$2348 & 2M 0407$-$2348  & 15.104 & 5 & $-$0.039 & 0.012 & $-$0.003 & 0.022 & 14.8 & 0.8 &  7.9 & 26.2 & 5.4 & M9.0 &  4.9 &  5    \\
D* 0419$-$4025 &                 & 18.900 & 5 &    0.040 & 0.018 & $-$0.020 & 0.011 & 15.6 & 1.1 &  8.0 & 23.6 & 4.3 & L1.0 &  5.0 &       \\
D* 0427$-$1127 &                 & 45.995 & 5 & $-$0.007 & 0.007 & $-$0.229 & 0.004 & 18.5 & 0.2 &  7.8 & 27.1 & 4.7 & M9.0 & 29.4 &       \\
D* 0436$-$4218 & 2M 0436$-$4218  &  5.987 & 4 &    0.033 & 0.031 &    0.028 & 0.036 & 16.2 & 2.5 &  8.1 & 24.5 & 4.7 & L2.0 &  5.0 &  7   \\
D* 0501$-$7705 &                 & 22.934 & 5 &    0.078 & 0.006 &    0.056 & 0.017 & 17.1 & 0.4 &  7.9 & 27.4 & 5.1 & L0.0 & 12.5 &       \\
D* 0526$-$4455 & 2M 0526$-$4455  & 19.187 & 4 &    0.030 & 0.007 & $-$0.161 & 0.008 & 18.7 & 0.3 &  8.2 & 18.7 & 3.4 & L2.0 & 14.5 &  2    \\
D* 0605$-$2342 & 2M 0605$-$2342  &  6.142 & 5 & $-$0.014 & 0.048 &    0.039 & 0.045 & 16.0 & 3.2 &  8.1 & 24.7 & 4.4 & L2.0 &  4.9 &  5    \\
D* 0624$-$4521 & 2M 0624$-$4521  &  5.663 & 3 & $-$0.040 & 0.031 &    0.236 & 0.027 & 20.2 & 0.5 &  8.4 & 16.2 & 3.1 & L4.0 & 18.4 &  3    \\
D* 0627$-$3919 &                 &  6.677 & 3 &    0.064 & 0.011 & $-$0.132 & 0.027 & 18.3 & 0.6 &  8.0 & 25.4 & 4.8 & L1.0 & 17.7 &       \\
D* 0641$-$4322 & S* 0641$-$4322  & 17.263 & 4 &    0.228 & 0.008 &    0.621 & 0.011 & 21.1 & 0.1 &  7.9 & 26.9 & 4.7 & L0.0 & 84.3 &  1    \\
D* 0703$-$6110 &                 &  6.964 & 3 &    0.070 & 0.009 &    0.005 & 0.024 & 17.1 & 0.5 &  8.0 & 28.4 & 5.4 & L1.0 &  9.4 &       \\
D* 0719$-$5051 &                 & 12.041 & 3 &    0.205 & 0.017 & $-$0.073 & 0.007 & 19.1 & 0.3 &  8.0 & 22.9 & 4.1 & L1.0 & 23.6 &       \\
D* 0921$-$2104 & S* 0921$-$2104  & 16.892 & 5 &    0.266 & 0.010 & $-$0.889 & 0.006 & 21.3 & 0.1 &  8.2 & 10.0 & 1.7 & L3.0 & 44.0 &  1,3  \\
D* 1019$-$2707 & S* 1019$-$2707  & 13.937 & 4 & $-$0.580 & 0.033 & $-$0.053 & 0.011 & 20.7 & 0.2 &  7.9 & 22.3 & 3.9 & L0.0 & 61.6 &  1,2  \\
D* 1115$-$2429 &                 &  8.017 & 4 & $-$0.001 & 0.024 & $-$0.198 & 0.042 & 18.0 & 0.5 &  7.8 & 25.8 & 4.4 & M9.0 & 24.2 &       \\
D* 1206$-$3937 &                 &  4.844 & 3 &    0.193 & 0.025 & $-$0.074 & 0.026 & 19.2 & 0.5 &  8.0 & 30.1 & 5.5 & L0.0 & 29.5 &       \\
D* 1234$-$1124 &                 & 12.977 & 4 & $-$0.137 & 0.023 & $-$0.104 & 0.020 & 19.4 & 0.6 &  8.2 & 24.6 & 4.9 & L2.0 & 20.1 &       \\
D* 1256$+$0146 &                 & 13.959 & 3 & $-$0.160 & 0.029 & $-$0.063 & 0.011 & 19.4 & 0.6 &  8.3 & 17.4 & 3.1 & L4.0 & 14.2 &       \\
D* 1359$-$4034 &                 &  5.961 & 3 &    0.066 & 0.011 & $-$0.531 & 0.010 & 20.6 & 0.1 &  7.9 & 28.4 & 5.2 & M9.0 & 72.0 &       \\
    \noalign{\smallskip}
    \hline 
   \end{tabular}
}
\end{table*}
\begin{table*}
\rotatebox{90}{
 \begin{tabular}{lcrcrlrlllllllrl}
\multicolumn{16}{p{22.5cm}}{{\bf Table 2.} Continued}\\
   \hline 
   \hline
   \noalign{\smallskip}
 DENIS name  &  Other name &  Time &  No. of   &$\mu$RA  &  err$\mu$RA &  $\mu$DE &  err$\mu$DE & 
 $H_{\rm I}$ &  err$H_{\rm I}$ &  $H_{\rm I}^{\rm max}$  & $d$         & err$d$   &  SpT  &  V$_{\rm t}$ &  Refs.  \\
             &             &  b.l. (yr) &  obs.  & (\arcsec yr$^{-1}$) &  (\arcsec yr$^{-1}$) &  (\arcsec yr$^{-1}$) & (\arcsec yr$^{-1}$) &
 (mag) &  (mag)  &  (mag) & (pc)  & (pc) &      &  km s$^{-1}$         &   \\
 (1) & (2) & (3) & (4) & (5) & (6) & (7) & (8) & (9) & (10) & (11) & (12) & (13) & (14) & (15) & (16) \\
   \hline
   \noalign{\smallskip}  
D* 1411$-$7915 &                 & 14.989 & 4 & $-$0.258 & 0.005 & $-$0.073 & 0.012 & 18.4 & 0.1 &  7.8 & 23.5 & 4.1 & M9.0 & 29.9 &       \\
D* 1446$-$3952 &                 & 17.095 & 3 &    0.061 & 0.004 & $-$0.007 & 0.005 & 16.6 & 0.3 &  8.0 & 29.0 & 5.9 & L1.0 &  8.4 &       \\
D* 1520$-$1755 &                 &  6.666 & 5 & $-$0.015 & 0.041 & $-$0.243 & 0.017 & 19.7 & 0.3 &  8.0 & 29.7 & 6.2 & L1.0 & 34.3 &       \\
D* 1622$-$1207 &                 &  6.762 & 3 &    0.020 & 0.024 & $-$0.056 & 0.009 & 15.4 & 0.7 &  7.9 & 23.8 & 4.2 & M9.0 &  6.7 &       \\
D* 1633$-$7553 &                 & 13.775 & 5 & $-$0.113 & 0.009 & $-$0.917 & 0.008 & 21.0 & 0.1 &  7.8 & 22.8 & 3.9 & M9.0 & 99.9 &       \\
D* 1707$-$0138 &                 & 12.115 & 4 &    0.070 & 0.031 & $-$0.023 & 0.017 & 17.1 & 1.2 &  8.1 & 22.0 & 4.3 & L2.0 &  7.7 &       \\
D* 1753$-$6559 & 2M 1753$-$6559  &  7.760 & 4 & $-$0.063 & 0.020 & $-$0.384 & 0.033 & 20.8 & 0.3 &  8.1 & 20.7 & 3.8 & L2.0 & 38.2 &  3    \\
D* 1901$-$3700$^{a}$ &           & 19.012 & 4 &    0.009 & 0.144 & $-$0.065 & 0.168 & 17.0 & 6.3 &  8.3 &      &     &      &      &  8    \\
D* 1907$-$2824 &                 & 11.064 & 3 &    0.091 & 0.014 & $-$0.206 & 0.006 & 19.7 & 0.3 &  8.2 & 21.8 & 4.2 & L2.0 & 23.3 &       \\
D* 1934$-$1841 &                 &  8.869 & 3 &    0.076 & 0.016 & $-$0.019 & 0.011 & 17.2 & 0.6 &  8.0 & 26.4 & 5.0 & L1.0 &  9.8 &       \\
D* 2013$-$1242 &                 &  3.937 & 3 &    0.025 & 0.077 & $-$0.067 & 0.044 & 17.3 & 2.2 &  8.1 & 24.8 & 5.1 & L2.0 &  8.4 &       \\
D* 2126$-$3143 &                 & 13.789 & 4 &    0.214 & 0.017 & $-$0.238 & 0.011 & 18.8 & 0.2 &  7.8 & 25.2 & 5.0 & M8.0 & 38.2 &       \\
D* 2139$-$3529 &                 & 16.019 & 3 &    0.012 & 0.010 & $-$0.059 & 0.006 & 16.8 & 0.5 &  8.1 & 26.7 & 5.0 & L1.0 &  7.6 &       \\
D* 2143$-$8337 &                 &  6.073 & 5 &    0.336 & 0.002 & $-$0.212 & 0.022 & 19.5 & 0.2 &  7.9 & 19.2 & 3.2 & L0.0 & 36.2 &  6    \\
D* 2150$-$6610 &                 & 10.998 & 4 & $-$0.073 & 0.007 &    0.079 & 0.015 & 17.4 & 0.5 &  8.1 & 16.5 & 2.9 & L2.0 &  8.4 &       \\
D* 2150$-$7520 & S* 2150$-$7520  &  6.820 & 3 &    0.839 & 0.001 & $-$0.198 & 0.009 & 22.1 & 0.2 &  8.1 & 20.1 & 3.6 & L2.0 & 82.1 &  1    \\
D* 2243$-$5932 & 2M 2243$-$5932  & 13.971 & 4 & $-$0.019 & 0.010 & $-$0.256 & 0.015 & 19.4 & 0.3 &  7.9 & 28.1 & 5.2 & L0.0 & 34.2 &  2    \\
D* 2308$-$2722 & S* 2308$-$2721  &  4.019 & 4 &    0.550 & 0.040 & $-$0.214 & 0.054 & 22.0 & 0.4 &  8.1 & 26.3 & 5.7 & L2.0 & 73.6 &  1    \\
D* 2345$+$0055 &                 &  9.880 & 4 &    0.050 & 0.040 & $-$0.050 & 0.020 & 16.1 & 1.4 &  7.9 & 28.3 & 5.6 & M9.0 &  9.5 &       \\
    \noalign{\smallskip}
    \hline\\ 
\multicolumn{16}{l}{Abbreviations.---D*: DENIS; 2M: 2MASS; S*: SIPS.}\\
\multicolumn{16}{l}{Cols. (1) \& (2): abbreviated DENIS name and other names.}\\
\multicolumn{16}{l}{Cols. (3) \& (4): number of observations and time baseline.}\\
\multicolumn{16}{l}{Cols. (5)--(8): proper motions and associated errors.}\\
\multicolumn{16}{l}{Cols. (9) \& (10): $I$-band reduced proper motion and associated error.}\\
\multicolumn{16}{l}{Col. (11): maximum reduced proper motion for an M giant of the same $I-J$ color.}\\
\multicolumn{16}{l}{Cols. (12) \& (13): photometric distance and its standard error.}\\
\multicolumn{16}{l}{Col. (14): spectral type.}\\
\multicolumn{16}{l}{Col. (15): tangential velocities.}\\
\multicolumn{16}{l}{References. ---(1): \citet*{deacon}; (2): \citet{kendall}; (3): \citet{reid06}; 
(4): \citet{bouy}; (5): \citet{cruz06};} \\
\multicolumn{16}{l}{(6): \citet{del99}; (7): \citet{reid08}; (8): \citet{martin10}.} \\
\multicolumn{16}{l}{$^{a}$: A member candidate of the
Corona Australis (R-CrA) molecular cloud complex \citep{martin10}, see \S~5 for further discussion.} \\
   \end{tabular}
}
\end{table*}
\begin{table*}
\setcounter{table}{2}
  \caption{2 red, distant objects}
\label{redobj}
  $$
 \begin{tabular}{lcrcrlrllll}
   \hline 
   \hline
   \noalign{\smallskip}
 DENIS name  &  Other name &  Time &  No. of   &$\mu$RA  &  err$\mu$RA &  $\mu$DE &  err$\mu$DE & 
 $H_{\rm I}$ &  err$H_{\rm I}$ &  $H_{\rm I}^{\rm max}$ \\
             &             &  b.l. (yr) &  obs.  & (\arcsec yr$^{-1}$) &  (\arcsec yr$^{-1}$) &  (\arcsec yr$^{-1}$) & (\arcsec yr$^{-1}$) &
 (mag) &  (mag)  &  (mag) \\
 (1) & (2) & (3) & (4) & (5) & (6) & (7) & (8) & (9) & (10) & (11) \\
   \hline
   \noalign{\smallskip}  
D* 1647$-$1156 &   &  3.270  & 3 &    0.021 & 0.074 & $-$0.006 & 0.041 & 14.8  & 8.4 &  9.0  \\
D* 1716$-$0315 &   & 46.800  & 5 & $-$0.005 & 0.006 &    0.001 & 0.003 &  8.0  & 2.8 &  8.0  \\
    \noalign{\smallskip}
    \hline 
   \end{tabular}
   $$
\end{table*}

\section{DISCUSSION}

Figure~\ref{HI_IJ} demonstrates that the final target
sample was only little ($\sim$3\%) contaminated 
by distant red objects. 
The large majority of distant red objects had been previously
rejected by the ($I-J$, $J-K$)
color-color filtering and the literature check. The
faint $I$-band magnitudes of the remaining candidates ($I \geq 16$) 
which could not be associated with a known giant
star made it unlikely that they could be unrecognized
ones: a giant with $I-J \geq 3.0$ ($\sim$M8) has
an $I$-band absolute magnitude  $M_{\rm I} \geq -3.0$ 
\citep{fluks}, which for $I \geq 16$ gives a 
minimum distance of $\sim$60~kpc, outside our 
Galaxy. Finally, our restriction to high Galactic 
latitudes ($|b_{\rm II}| \geq 15\degr$) and our 
rejection of objects which overlap known molecular 
complexes (mostly) eliminated reddened bright
stars as contaminants. Within those restrictions,
and for application where a small residual contamination 
is acceptable, one could thus in principle dispense with 
the MRPM filtering. 

Using our $M_{\rm J}$ vs. $I-J$ calibration, we estimated 
distances for all 57 ultracool dwarfs, and then computed 
tangential velocities, (see Table~2). One should note that
Table~2 does not list these estimates for DENIS~1901$-$3700,
a member candidate of the R-CrA molecular cloud region \citep{martin10} 
(see below). 
Four of the dwarfs have $V_{\rm t}$ above 80 km s$^{-1}$ 
(DENIS 0014$-$4844, DENIS 0641$-$4322, DENIS 1633$-$7553, 
and DENIS 2150$-$7520), and could thus have reduced 
metallicities. All ultracool subdwarfs 
($I-J~\geq~2.0$, later than $\sim$sdM9) known to
date however have $J-K<0.7$ (e.g., Table~7 
in \citealt*{bur06} and references therein). The
$J-K \geq 0.89$ of these four stars therefore 
make it unlikely that they are true subdwarfs,
they could still have somewhat subsolar
metallicities.

One should note that some of the 57 ultracool dwarfs 
could be unresolved binaries, whose distances are 
underestimated by up to $\sqrt{2}$. \citet{reid06}
recently reported that DENIS~0147$-$4954 (as 2MASS~0147$-$4954) 
is an M8+L2 binary.

\begin{figure}
\psfig{width=8.2cm,file=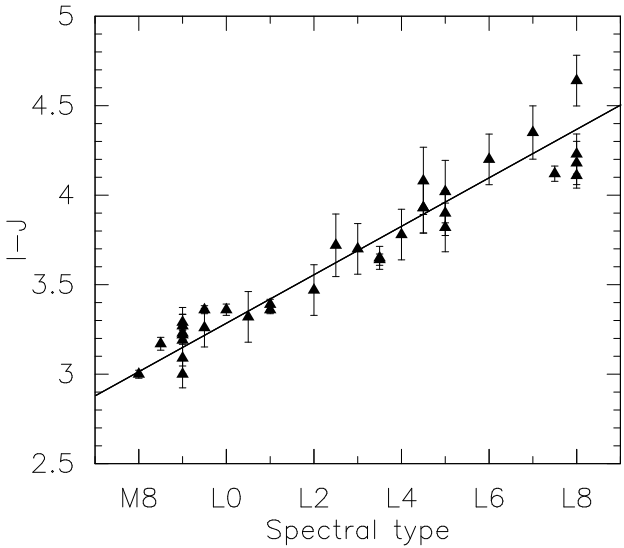,angle=-90}
\caption{Color-spectral type diagram for 34 late-M 
($\geq$M8.0) and L dwarfs, plotted as solid triangles (data in Table~1 of
\citealt{p08}).
\label{IJ_SpT}}
\end{figure}

\begin{figure}
\psfig{width=8.2cm,file=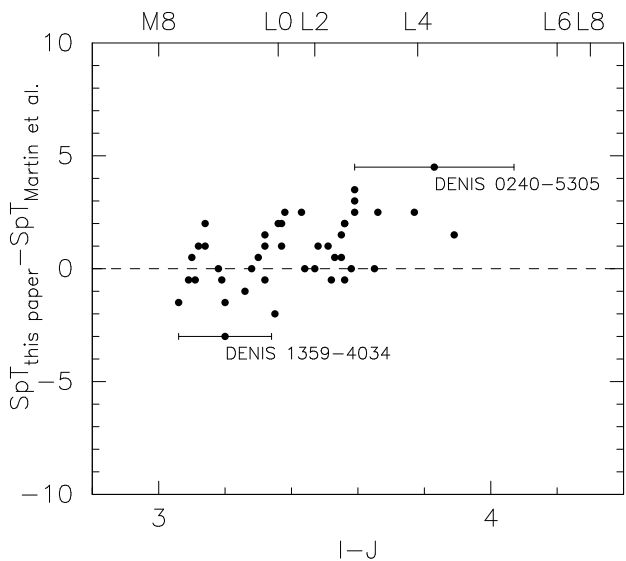,angle=-90}
\caption{Comparison of our photometrically estimated spectral types
of 43 ultracool dwarfs with the spectroscopically determined
ones of \citet{martin10}.
Two objects DENIS 0240$-$5305 and DENIS 1359$-$4034, which have large
deviations, are indicated with error bars on the $I-J$ color.
\label{SpT_comparison}}
\end{figure}

We estimate spectral types for the 57 ultracool dwarfs
from their $I-J$ color. A linear least-squares fit (Fig.~\ref{IJ_SpT})
to the data of 34 late-M, and L dwarfs
(Table~1 in \citealt{p08}) gives the following color-spectral type relation: 
SpT~$= 7.39(I-J) - 14.28$, $\sigma=0.87$, where SpT is the spectral subtype,
counted from 8.0 for spectral type M8.0 to 18.0 for spectral type L8.0.
Table~2 lists the resulting spectral types for the 57 ultracool dwarfs,
with a standard error of about 1~subclass.
Figure~\ref{SpT_comparison} shows a comparison of our estimated
spectral types of 43 ultracool dwarfs with the spectroscopically
determined ones of \citet{martin10}. Some objects have
large deviations, e.g., DENIS 0240$-$5305 and DENIS 1359$-$4034,
clearly due to large photometric uncertainties.

DENIS 1901$-$3700 is above our MRPM curve within 2$\sigma$.
The large relative error on its proper 
motion makes its true position in the 
($H_{\rm I}$, $I-J$) diagram uncertain,
reddening probably contributes to that
situation as the object lies just outside our 
mask for the R Coronae Australis molecular cloud.
The extremely red color of DENIS 1901$-$3700 
($I-J=3.71$) likely reflects a combination
of reddening with a red intrinsic spectrum.
\citet{martin10} reported that DENIS 1901$-$3700
has a low surface gravity and M8 spectral type and is
a member candidate of the R-CrA region. However
its spectrum does not show strong H$\alpha$ emission
as seen in DENIS-P J1859509$-$370632 \citep{bouy04}.
Further observations are required to confirm
its membership in R-CrA or its status as a background star.

DENIS 1647$-$1156 is above the MRPM curve within 1$\sigma$
and it is a reddened main sequence star (M. Bessell, private communication).
As the star is in the general vicinity of the 
LDN 204B dark nebula, it is probably reddened 
by dust associated with that region. 
DENIS 1716$-$0315 lies on the curve and the source
is a giant star \citep{martin10}.
This results in a fraction of about 3\% contaminated by red, distant
stars in our sample of nearby ultracool dwarfs.

\section{SUMMARY}
In this paper, we have shown 
how to use the MRPM method to detect
57 nearby L and late-M dwarfs in the DENIS database. 
We measured proper motions of these ultracool dwarfs: 
38 with high-pm and 19 with low-pm,
giving a significant $\sim$33\% fraction of 
low-pm ultracool dwarfs in our sample.
These sources
are in the solar neighborhood and cooler than M8. 
They are therefore new targets for a search of planets
around ultracool dwarfs. Their detection demonstrates
the success of the MRPM method that can be used in the future to complete
the search of nearby ultracool dwarfs 
in the Galactic plane (the northern hemisphere part).

\section*{Acknowledgments}
The DENIS project has been partly funded by the SCIENCE and the
HCM plans of the European Commission under grants CT920791
and CT940627. It is supported by INSU, MENand CNRS in France,
by the State of Baden-W\"urttemberg in Germany, by DGICYT in
Spain, by CNR in Italy, by FFwFBWF in Austria, by FAPESP in
Brazil, by OTKA grants F-4239 and F-013990 in Hungary, and by
the ESO C\&EE grant A-04-046. Jean Claude Renault from IAP
was the project manager. Observations were carried out thanks to
the contribution of numerous students and young scientists from
all involved institutes, under the supervision of P. Fouqu\'e, survey
astronomer resident in Chile.

The Digitized Sky Surveys were produced at the Space 
Telescope Science Institute under U.S. Government 
grant NAG W-2166. 
The images of these surveys are based on photographic 
data obtained using the Oschin Schmidt Telescope 
on Palomar Mountain and the UK Schmidt Telescope. 
The plates were processed into the present compressed 
digital form with the permission of these institutions. 
The National Geographic Society - Palomar Observatory 
Sky Atlas (POSS-I) was made by the California Institute 
of Technology with grants from the National Geographic Society. 
The Second Palomar Observatory Sky Survey (POSS-II) was 
made by the California Institute of Technology with 
funds from the National Science Foundation, the 
National Geographic Society, the Sloan Foundation, 
the Samuel Oschin Foundation, and the Eastman Kodak 
Corporation. 
The Oschin Schmidt Telescope is operated 
by the California Institute of Technology and Palomar 
Observatory. The UK Schmidt Telescope was operated by 
the Royal Observatory Edinburgh, with funding from the 
UK Science and Engineering Research Council (later the 
UK Particle Physics and Astronomy Research Council), 
until 1988 June, and thereafter by the Anglo-Australian 
Observatory. The blue plates of the southern Sky Atlas 
and its Equatorial Extension (together known as the SERC-J), 
as well as the Equatorial Red (ER), and the Second Epoch 
[red] Survey (SES) were all taken with the UK Schmidt.
This publication makes use of data products from
the Two Micron All Sky Survey, which is a joint project 
of the University of Massachusetts and Infrared
Processing and Analysis Center/California Institute of 
Technology, funded by the National Aeronautics
and Space Administration and the National Science 
Foundation; the NASA/IPAC Infrared Science Archive, which
is operated by the Jet Propulsion Laboratory/California 
Institute of Technology, under contract with the 
National Aeronautics and Space Administration.
This research has made use of the ALADIN, SIMBAD and VIZIER databases, 
operated at CDS, Strasbourg, France. 

N.P.-B has been supported by Vietnam National University HCMC 
grant B2011-28-10.

\label{lastpage}
\end{document}